\documentclass[12pt]{article}

\usepackage{amsmath}
\usepackage{amssymb}

\newcommand{\al}{\alpha}
\newcommand{\ep}{\epsilon}

\newcommand{\del}{\partial}

\newcommand{\bm}[1]{\mbox{\boldmath ${#1}$}}

\newcommand{\msc}[1]{\mbox{\scriptsize #1}}
\newcommand{\dsp}{\displaystyle}

\newcommand{\bc}{\Bbb C}
\newcommand{\br}{\Bbb R}
\newcommand{\bz}{\Bbb Z}

\newcommand{\bh}{\Bbb H}

\newcommand{\bsz}{\Bbb Z}

\newcommand{\f}{\mbox{{\bf f}}}

\newcommand{\cJ}{{\cal J}}
\newcommand{\cT}{{\cal T}}

\newcommand{\cN}{{\cal N}}

\newcommand{\tL}{\tilde{L}}

\newcommand{\tF}{\widetilde{F}}

\newcommand{\tep}{\tilde{\epsilon}}

\newcommand{\ts}{\tilde{s}}

\newcommand{\ket}[1]{{\left|#1\right\rangle}}

\newcommand{\Th}[2]{\Theta_{#1,#2}}
\renewcommand{\th}{{\theta}}
\newcommand{\tTh}[2]{\widetilde{\Theta}_{#1,#2}}

\newcommand{\tr}{\mbox{Tr}}

\newcommand{\nn}{\nonumber\\}

\newcommand{\NS}{\mbox{NS}}
\newcommand{\tNS}{\widetilde{\mbox{NS}}}
\newcommand{\R}{\mbox{R}}

\newcommand{\sNS}{\msc{NS}}

\newcommand{\sR}{\msc{R}}

\newcommand{\any}{{}^{\forall}}

\usepackage{color}
\def\red#1  {{\textcolor{red}{#1 }}}

\newcommand {\eqn}[1]{(\ref{#1})}

\makeatletter
\@addtoreset{equation}{section}
\def\theequation{\thesection.\arabic{equation}}
\makeatother

\begin{document}


\begin{titlepage}
 \
 \renewcommand{\thefootnote}{\fnsymbol{footnote}}
 \font\csc=cmcsc10 scaled\magstep1
 {\baselineskip=16pt
  \hfill
 \vbox{\hbox{December, 2015}
 \hbox{UTHEP-679}
       }}

 \baselineskip=20pt
\vskip 1cm
 
\begin{center}

{\bf \Large

Non-supersymmetric Asymmetric Orbifolds

with 

Vanishing Cosmological Constant 

} 

 \vskip 1.2cm

\noindent{ \large Yuji Satoh}\footnote{\sf ysatoh@het.ph.tsukuba.ac.jp}
 \\

\medskip

{\it Institute of Physics, University of Tsukuba, \\
 Ibaraki 305-8571,  Japan}

\vskip 8mm
\noindent{ \large Yuji Sugawara}\footnote{\sf ysugawa@se.ritsumei.ac.jp},
\hspace{1cm}
\noindent{ \large Taiki Wada}\footnote{\sf rp0017xp@ed.ritsumei.ac.jp},

\medskip

 {\it Department of Physical Sciences, 
 College of Science and Engineering, \\ 
Ritsumeikan University,  
Shiga 525-8577, Japan}

\end{center}

\bigskip

\begin{abstract}

We study type II string vacua defined by  torus  
compactifications accompanied by T-duality twists.
We realize the string vacua, specifically,  by means of 
the asymmetric orbifolding 
associated to the chiral reflections combined with a shift,
which are interpreted as describing the compactification on `T-folds'.
We discuss possible consistent actions of the chiral reflection on 
the Ramond-sector of the world-sheet fermions, 
and explicitly construct non-supersymmetric as well as supersymmetric vacua. 
Above all, we demonstrate a simple realization of non-supersymmetric vacua 
with vanishing cosmological constant at one loop.  
Our orbifold group is generated  only by a single element,  
which results in simpler models than those with such property known previously.

\end{abstract}

\setcounter{footnote}{0}
\renewcommand{\thefootnote}{\arabic{footnote}}

\end{titlepage}

\baselineskip 18pt

\vskip2cm 
\newpage


\section{Introduction}

Compactifications on non-geometric backgrounds have been receiving  
increasing attention in superstring theory. A particularly interesting class of 
such backgrounds  
is formulated as the fibrations of which  the transition functions 
involve the duality transformations 
in string theory 
\cite{Dabholkar:2002sy,Hellerman:2002ax,Flournoy:2004vn}.  
For T-duality, for instance, one then has `T-folds' \cite{Hull:2004in}.
Another interesting class is the backgrounds with non-geometric fluxes that 
do not have naive geometrical origins in  higher dimensional theories. 
In some cases, these are reduced to geometric ones by dualities, but 
are  truly non-geometric in general 
\cite{Shelton:2005cf,Shelton:2006fd,Dabholkar:2005ve}.

These string vacua on non-geometric backgrounds are described 
by the world-sheet conformal field theory (CFT) on the same footing
as geometric ones. 
We should emphasize that many of such vacua are well-defined only 
at particular points 
on the moduli space, at which enhanced symmetries emerge 
and the $\al'$-corrections become important. 
The world-sheet CFT approach would provide reliable descriptions of strings 
even in such backgrounds. In this respect, 
a simple and important class of non-geometric backgrounds is realized as 
asymmetric orbifolds \cite{Narain:1986qm}, in which the left- and the right-movers 
of strings propagate on different geometries. 
Especially, as typical T-duality twists are identified with chiral reflections, 
simple examples of  T-folds are realized as the orbifolds 
by the chiral reflection combined with the shift in the base circle. 
These types of string vacua have been 
studied based on the world-sheet CFT  {\em e.g.\/}  in
 \cite{Flournoy:2005xe,HW,KawaiS1,KawaiS2,
 Condeescu:2012sp,Condeescu:2013yma,SatohS,Tan:2015nja}.%
\footnote{
For the aspects of non-commutativity in non-geometric backgrounds,
see {\em e.g.\/} \cite{Blumenhagen:2010hj,Lust:2010iy,
Blumenhagen:2011ph,Condeescu:2012sp,Andriot:2012vb}
}


In this paper, we study type II string vacua defined by torus 
compactifications twisted by T-duality transformations
in the above sense. 
We carefully discuss possible consistent actions of the chiral reflection 
on the Ramond sector of the world-sheet fermions, and explicitly construct 
non-supersymmetric as well as supersymmetric (SUSY) vacua.\footnote
  {For non-supersymmetric orbifolds in heterotic string theory, see 
{\em e.g.}
\cite{Blaszczyk:2014qoa,Angelantonj:2014dia,Faraggi:2014eoa,Abel:2015oxa} and references therein.} 
Among others, we present 
a simple realization of non-SUSY vacua 
{\em with vanishing cosmological constant\/} 
at the one-loop level, at least. 
Namely, we construct the string vacua realizing the bose-fermi cancellation 
despite the absence of any supercharges in space-time.
Previous constructions of such string vacua  are found
{\em e.g.} in \cite{Kachru1,Kachru2,Kachru3,Harvey,Shiu-Tye,Blumenhagen:1998uf,
Angelantonj:1999gm}.\footnote
{In the papers \cite{Kachru1,Kachru2,Kachru3}, the authors further conjectured 
 that the cosmological constant remains vanishing at 
 two and higher loops. However, 
a careful world-sheet analysis \cite{ADP} shows that it  does not actually
vanish at two loops in those models, 
at least pointwise on the moduli space.
} 
A novel feature, as well as an advantage, in our construction is
that we only have  to utilize 
a cyclic orbifold, in which the orbifold group is generated by a single element,  and hence
the construction looks rather simpler than 
previous ones given in those papers.  
It would be notable that one can achieve (nearly) vanishing cosmological constant
without SUSY in a fairly simple way 
in the framework of non-geometric string compactifications. 
Our construction suggests that they would provide useful grounds also for 
the cosmological constant problem.

To be more precise,  we first analyze in some detail the asymmetric orbifolds  
representing T-folds, where the partition sums from each sector 
in the total partition function
are combined according to the windings around the `base' circle.
It turns out that the consistent action of the chiral reflections therein 
is not unique, from which a variety of supersymmetric T-fold vacua can be derived. 
As general for asymmetric orbifolds, the moduli of the internal (`fiber') tori are fixed
for consistency, while a continuous radius of the base circle remains.
The supersymmetry is broken by further implementing the Scherk-Schwarz 
type boundary condition  for the world-sheet fermions \cite{SS1,SS2} 
along the base circle. 
In the case where the  chiral reflections act as $\bz_4$ transformations 
in a fermionic sector, the resultant 
world-sheet torus partition function and hence the one-loop 
cosmological constant vanish: if the partition sum for the left-moving fermions 
is non-vanishing in a winding sector, that for the right-moving fermions vanishes,
and vice versa.  It is crucial here that the chiral partition sums for the fermions 
depend on the winding numbers in an asymmetric way.
We see that
all the  ingredients in our setup, i.e., T-folds (asymmetric orbifolds, base winding),
careful treatment of the chiral reflections and the Scherk-Schwarz twist, 
cooperate in this mechanism. Although we focus on specific examples in this paper, 
our construction would be  more general. It provides a systematic way to 
find string vacua of T-folds, and a novel mechanism for 
non-supersymmetric string vacua with vanishing cosmological constant at one-loop.

~


This paper is organized as follows:
In section 2, which is a preliminary section,  
we survey the building blocks (partition sums) for the modular invariant partition 
functions of the asymmetric orbifolds discussed later, 
specifying how to achieve the modular covariance in relevant sectors.
Though this part might be slightly technical, the results, especially those 
for the fermionic sector presented in subsection \ref{fermionic sector},
are important in the later discussion both on T-fold vacua
and on vanishing cosmological constant.
The readers may refer only to the definitions of the building blocks, 
if they are interested mostly in the physical consequences. 


In section 3, we begin our main analysis of type II string vacua 
compactified on asymmetric orbifolds/T-folds. 
We first consider the supersymmetric ones. 
The SUSY breaking is then discussed 
by  further incorporating  the Scherk-Schwarz twist,
which leads us to the non-SUSY vacua implementing the bose-fermi cancellation. 
In section 4, we analyze the spectra of the physical states and check the unitarity, 
mainly focusing 
on the case of the non-SUSY vacua. We also demonstrate 
the absence of the instability caused by the winding tachyons, which would 
 be typically possible   for the Scherk-Schwarz compactification.
We conclude with a summary and 
a discussion for possible future directions in section 5.

~


\section{Preliminaries: Building Blocks for Asymmetric Orbifolds}

In this paper, we would like to study the type II string vacua 
constructed from asymmetric orbifolds of the 
10-dimensional flat background given by
\begin{equation}
M^4
\times S^1 \times \br_{\msc{base}} \times T^4_{\msc{fiber}},
\label{setup}
\end{equation}
where $M^4$ ($X^{0,1,2,3}$-directions) is the 4-dimensional  Minkowski space-time.
Intending the twisted compactification of the `base space' $\br_{\msc{base}}$
 ($X^5$-direction),
we consider the orbifolding defined by the twist operator
$
 \cT_{2\pi R} \otimes \sigma :
$
$\cT_{2\pi R}$ is the translation along the base direction by $2\pi R$,
and $\sigma$ denotes an automorphism acting on the `fiber sector' 
$T^4_{\msc{fiber}}$ ($X^{6,7,8,9}$), which is specified in detail later.
We especially focus on the cases where $\sigma$
acts as the `chiral reflection', or the T-duality transformation,
\begin{equation}
\label{sigma T4}
\sigma
~ : ~ (X_L^i, X_R^i)~ \longmapsto ~ (X_L^i, -X_R^i), ~~~ (i=6,7,8,9).
\end{equation}
The $S^1$-factor ($X^4$-direction)
in \eqn{setup} is not important in our arguments.
We begin our analysis by specifying the relevant bosonic and fermionic sectors 
 and their chiral blocks 
that compose the modular invariants for our asymmetric orbifolds.

\subsection{Bosonic $T^4_{\msc{fiber}}$ Sector}
\label{bosonic sector}

In the bosonic sector, 
let us first  consider the 4-dimensional 
torus with  the $SO(8)$-symmetry enhancement
which we denote as $T^4[SO(8)]$, 
in order that the relevant asymmetric orbifold action (chiral reflection) is well-defined.
The torus partition function of this system is 
\begin{eqnarray}
 && Z_{T^4[SO(8)]}(\tau,\bar{\tau}) 
= \frac{1}{2}\left\{
\left|\frac{\th_3}{\eta}\right|^8+\left|\frac{\th_4}{\eta}\right|^8
+\left|\frac{\th_2}{\eta}\right|^8
\right\}. 
\label{Z T4 SO(8)}
\end{eqnarray}
Another system 
that is compatible with our asymmetric orbifolding and of our interests
is the product of  the 2-dimensional 
tori with the $SO(4)$-symmetry,
$
T^2[SO(4)] \times T^2[SO(4)] ,
$ 
the partition function of which is given by 
\begin{equation}
Z_{T^2[SO(4)] \times T^2[SO(4)]] }(\tau,\bar{\tau}) 
= \frac{1}{4}\left\{
\left|\frac{\th_3}{\eta}\right|^4+\left|\frac{\th_4}{\eta}\right|^4
+\left|\frac{\th_2}{\eta}\right|^4
\right\}^2.
\label{Z T4 2 SO(4)}
\end{equation}
It is useful to note the equivalence
\begin{equation}
T^2[SO(4)] \times T^2[SO(4)]
\cong T^4[SO(8)]/ \bz_2
\cong 
\left[ S^1 [SU(2)] \right]^4,
\label{T2 T2 rel}
\end{equation}
where $S^1 [SU(2)] $ expresses the circle of the self-dual radius $R=1$.\footnote
 {Throughout this paper, we 
 use the $\al'=1$ convention.} 
Namely, 
while both  of $X^{6,7}$ and $X^{8,9}$ are compactified on 
the 2-torus $T^2[SO(4)]$ at the fermionic point with radius $\sqrt{2}$, 
the following four compact bosons have the self-dual radius, 
\begin{equation}
Y^1_{\pm} := \frac{1}{\sqrt{2}} \left(X^6 \pm i X^7\right) , 
\hspace{1cm}
Y^2_{\pm} := \frac{1}{\sqrt{2}} \left(X^8 \pm i X^9\right) .
\label{def Ypm}
\end{equation} 
The equivalence \eqn{T2 T2 rel} is confirmed by the simple identities
\eqn{Z T4 2 SO(4) app}.



We then consider the action of the automorphism $\sigma$  for $T^4[SO(8)] $ and 
$T^2[SO(4)] \times T^2[SO(4)]$.
 Since relative phases for the left and the right movers are generally possible
in asymmetric orbifolding,
in addition to the action without phases, 
we consider an action with phases according to \cite{HW} for 
$T^2[SO(4)] \times T^2[SO(4)]$.
In total, 
we consider the following three cases as models relevant to our construction of string vacua given in section 3. 
This means that the moduli of $T^4_{\msc{fiber}}$ need be restricted to the particular points given here,
while the radius of $S^1_{\msc{base}}$ can be freely chosen.   
We particularly elaborate on the derivation of the building blocks for the case of $T^4[SO(8)] $,
and mention on other cases briefly. The explicit forms of  
the relevant building blocks are summarized in Appendix B. 
The case with phases for $T^4[SO(8)] $  can 
be similarly discussed following \cite{HW,Tan:2015nja}, although we do not work on it in this paper.

~

\noindent
{\bf 1. \ Chiral reflection in {$T^4[SO(8)]$}}

\par\bigskip
We start with $T^4[SO(8)]$. 
In this case, the orbifold action is 
defined by the  chiral reflection \eqn{sigma T4} 
acting only on the right-moving components. 
We simply assume 
$\sigma$ acts as the identity operator on any states in 
the left-mover, and also that $\sigma^2$
acts  as the identity over the Hilbert space of 
the {\em untwisted\/} sector of the orbifolds of our interest.\footnote
   {This assumption is not necessarily obvious. Actually, if we fermionize 
   the string coordinates 
along $T^4[SO(8)]$, we can also realize more general situations 
as in our discussion given in subsection \ref{fermionic sector}.
We do not study  these cases for simplicity in this paper. 
}
We note that the action of $\sigma^2$
on the  {\em twisted\/} sectors  should be determined 
so that it preserves the modular invariance of the total system, and does 
not necessarily coincide with the identity. 
This is a characteristic feature of 
asymmetric orbifolds.  
See for example \cite{Aoki:2004sm}.

Let us evaluate the building blocks in this sector of the torus partition function.
These are schematically written as 
\begin{equation}
F^{T^4[SO(8)]}_{(a,b)}(\tau,\bar{\tau}) := \tr_{\msc{$\sigma^a$-twisted sector}} 
\left[\sigma^b \, 
q^{L_0-\frac{c}{24}}\bar{q}^{\tL_0-\frac{c}{24}}\right].
\label{F T4 ab def}
\end{equation}
Here, we allow $a$, $b$ to be any integers despite 
a periodicity, which is at most of order 4 as seen below, since we 
later identify  them as the winding numbers along the base circle $S^1_{\msc{base}}$. 
We can most easily determine the building blocks $F^{T^4[SO(8)]}_{(a,b)}$ by requiring 
the  modular covariance,
\begin{eqnarray}
 && F^{T^4[SO(8)]}_{(a,b)}(\tau,\bar{\tau})|_S = F^{T^4[SO(8)]}_{(b,-a)}(\tau,\bar{\tau}), \nn
 && F^{T^4[SO(8)]}_{(a,b)}(\tau,\bar{\tau})|_T = F^{T^4[SO(8)]}_{(a,a+b)}(\tau,\bar{\tau}),
\label{T4 SO(8) mod cov} 
\end{eqnarray}
together with  the trace over the untwisted sector, 
\begin{equation}
F^{T^4[SO(8)]}_{(0,b)}(\tau,\bar{\tau}) =  
\overline{\left(\frac{\th_3\th_4}{\eta^2}\right)^2} \cdot 
\frac{1}{2} \left\{
\left(\frac{\th_3}{\eta}\right)^4
+ \left(\frac{\th_4}{\eta}\right)^4
\right\}, 
\hspace{1cm} 
(\any b \in 2\bz+1).
\end{equation}
Then, the desired building blocks are found to be 
\begin{eqnarray}
&& 
\hspace{-1cm}
F^{T^4[SO(8)]}_{(a,b)}(\tau,\bar{\tau}) =
\left\{
\begin{array}{ll}
(-1)^{\frac{a}{2}}
 \overline{\left(\frac{\th_3\th_4}{\eta^2}\right)^2}
\cdot 
\frac{1}{2}\left\{
\left(\frac{\th_3}{\eta}\right)^4
+\left(
\frac{\th_4}{\eta}
\right)^4
\right\}
& ~~ (a\in 2\bz, ~ b\in 2\bz+1), \\
(-1)^{\frac{b}{2}}
 \overline{\left(\frac{\th_2\th_3}{\eta^2}\right)^2}
\cdot 
\frac{1}{2}\left\{
\left(\frac{\th_3}{\eta}\right)^4+\left(
\frac{\th_2}{\eta}
\right)^4
\right\}
& ~~ (a\in 2\bz+1, ~ b\in 2\bz), \\
 e^{-\frac{i\pi}{2}ab} 
 \overline{\left(\frac{\th_4\th_2}{\eta^2}\right)^2}
\cdot 
\frac{1}{2}\left\{
\left(\frac{\th_4}{\eta}\right)^4- \left(
\frac{\th_2}{\eta}
\right)^4
\right\}
& ~~ (a\in 2\bz+1, ~ b\in 2\bz+1), \\
\frac{1}{2}\left\{
\left|\frac{\th_3}{\eta}\right|^8 + \left|\frac{\th_4}{\eta}\right|^8
+\left|\frac{\th_2}{\eta}\right|^8
\right\}
& ~~ (a \in 2\bz, ~ b\in 2\bz) .
\end{array}
\right. 
\label{Fab T4}
\end{eqnarray}
One can 
confirm that they indeed satisfy the modular covariance relations
\eqn{T4 SO(8) mod cov}.

~


\noindent
{\bf 2. \ Chiral reflection in {$T^2[SO(4)] \times T^2[SO(4)]$}}

\par\bigskip

In the first case of 
$T^2[SO(4)] \times T^2[SO(4)]$ or the $\bz_2$-orbifold of $T^4[SO(8)]$,
we may consider the same orbifold action $\sigma$ as given in  case 1, 
Namely, it acts as the identity on the left-mover, and assumes $\sigma^2 = \bm{1}$ in the untwisted sector. 
The modular covariant building blocks of the torus partition function are just
determined in the same way as above. We present them in \eqn{Fab T2 T2} in Appendix B.

~


\noindent
{\bf 3. \ Chiral reflection in {$T^2[SO(4)] \times T^2[SO(4)]$} with a phase factor}

\par\bigskip

In the second case of $T^2[SO(4)] \times T^2[SO(4)]$,
we include the phase factors for the Fock vacua 
when defining $\sigma$,  
while the action of the chiral reflection \eqn{sigma T4} is kept unchanged.
To be more specific, recalling the equivalence \eqn{T2 T2 rel}, let us  
introduce 4 copies of the $SU(2)$-current algebra of level 1 whose 
third components  are identified as 
\begin{equation}
J^{3, \, (i)} = i\del Y_+^1, ~ i\del Y_-^1, ~ i\del Y_+^2, ~ i\del Y_-^2, 
\hspace{1cm} (i=1,\ldots, 4),
\end{equation}
where $Y$'s are the compact bosons in \eqn{def Ypm}.
With these currents, $\sigma$ is now explicitly defined according to \cite{HW}  by
\begin{equation}
\sigma
:=\prod_{i=1}^4 \, \left[ 
 e^{i\pi J^{3, \, (i)}_{L, 0}} \otimes e^{i \pi J_{R, 0}^{1, \, (i)} }
\right].
\label{sigma T2 T2 2}
\end{equation}
We then obtain the building blocks according to the same procedure :  
the blocks for the $(0, b)$-sectors with $\any b \in 2\bz+1$ are computed first, and then those for 
other sectors are obtained by requiring the modular covariance. 
It turns out that these are eventually
equal to the building blocks of the symmetric $\bz_2$-orbifold 
defined by
\begin{equation}
(X^i_L, X^i_R) ~ \longmapsto ~ (-X^i_L, -X^i_R), \hspace{1cm} (\any i =6,7,8,9).
\end{equation}
Of course, this fact is not surprising since 
\eqn{sigma T2 T2 2} is equivalent to 
the symmetric one
\begin{equation}
\prod_{i=1}^4 \, \left[ 
 e^{i\pi J^{1, \, (i)}_{L, 0}} \otimes e^{i \pi J_{R, 0}^{1, \, (i)} }
\right],
\end{equation}
by an automorphism of $SU(2)^{\otimes 4}$, 
as was pointed out in \cite{KawaiS1}.
We exhibit the  building blocks in this case 
in \eqn{tFab T2 T2}.


\subsection{Fermionic Sector}
\label{fermionic sector}

We next consider the fermionic sector. 
The orbifold action should act on the world-sheet fermions as 
\begin{equation}
\sigma
 ~:~ (\psi_L^i, \psi^i_R) ~ \longmapsto 
~ (\psi_L^i, - \psi^i_R), \hspace{1cm} (i=6,7,8,9),
\label{sigma fermion}
\end{equation}
to preserve the world-sheet superconformal symmetry. 
 \eqn{sigma fermion} uniquely determines the action on the Hilbert space of 
the NS-sector. 
However, it is not  on the R-sector, and 
as is discussed in the next section, we obtain different 
string vacua according to its choice.
The fermionic part is thus crucial in our analysis.
In the following, we include the fermions $\psi^i$ $(i=2,3,4,5)$ in other transverse 
part from 
$M^4 \times S^1 \times S^1_{\rm base}$, on which $\sigma$ acts trivially.
If retaining the Poincare symmetry in 4 dimensions,
we then have two possibilities, which can be
understood from the point of view of bosonization
as follows: 


~

\noindent
{\bf (i) \  $\bz_2$ action on the untwisted R-sector}
\par\bigskip

In this case, we bosonize $\psi^i_R$ ($i=2, \ldots, 9$) as
\begin{eqnarray}
&& \psi^2_R \pm i \psi^3_R \equiv \sqrt{2} e^{\pm iH_{0, R}},
\qquad 
\psi^4_R \pm i \psi^5_R \equiv \sqrt{2} e^{\pm iH_{1, R}}, \nonumber \\
&& \psi^6_R \pm i \psi^7_R \equiv \sqrt{2} e^{\pm iH_{2, R}},
\qquad 
\psi^8_R \pm i \psi^9_R \equiv \sqrt{2} e^{\pm iH_{3, R}},
\end{eqnarray}
and define the spin fields for $SO(8)$ as 
\begin{equation}
S_{\ep_0, \ep_1 \ep_2 \ep_3, R} \equiv e^{\frac{i}{2} \sum_{i=0}^3 
\ep_i H_{i,R}}, \hspace{1cm} (\ep_i = \pm 1) .
\label{Spin 1}
\end{equation}
Then, \eqn{sigma fermion}  translates into 
\begin{equation}
\sigma
~: ~  
 (H_{0, R}, H_{1, R}, H_{2,R}, H_{3,R} ) ~ \longmapsto ~ 
 (H_{0, R}, H_{1, R}, H_{2,R} + \pi , H_{3,R} + \pi),
\label{sigma H}
\end{equation}
and thus, we find 
$\sigma^2 = \bm{1}$ for all the states in the NS and R-sectors 
in the untwisted sector. 
This type of twisting preserves half of the space-time SUSY. 
In fact, the Ramond vacua that are generated by 
the spin fields \eqn{Spin 1} 
survive the $\sigma$-projection when $\ep_2+ \ep_3=0$.


~

\noindent
{\bf (ii) \  $\bz_4$ action on the untwisted  R-sector}
\par\bigskip

 In this case, we bosonize $\psi^i_R$ ($i=2, \ldots, 9$) as 
\begin{eqnarray}
&& H'_{0, R} \equiv H_{0, R} , \hspace*{15ex}
 \psi^4_R \pm i \psi^6_R \equiv \sqrt{2} e^{\pm iH'_{1, R}}, \nonumber \\
&&
\psi^5_R \pm i \psi^7_R \equiv \sqrt{2} e^{\pm iH'_{2, R}},
\qquad 
 H'_{3, R} \equiv H_{3, R} ,
\end{eqnarray}
and define the spin fields for $SO(8)$ as 
\begin{equation}
S'_{\ep_0, \ep_1 \ep_2 \ep_3, R} \equiv e^{\frac{i}{2} 
\sum_{i=0}^3 
\ep_i H'_{i,R}}, \hspace{1cm} (\ep_i = \pm 1).
\label{Spin 2}
\end{equation}
This time,  
\eqn{sigma fermion}  translates into 
\begin{equation}
\sigma
~: ~ 
 (H'_{0, R},H'_{1, R}, H'_{2,R}, H'_{3,R} ) ~ \longmapsto ~ 
 (H'_{0, R}, -H'_{1, R}, -H'_{2,R} , H'_{3,R} + \pi).
\label{sigma H'}
\end{equation}
Then, 
 $\sigma^2 = - \bm{1}$ for the R-sector, while $\sigma^2 = \bm{1}$ still holds 
for the NS sector. 
In other words, we have found in this second case that 
\begin{equation}
\sigma^2 = (-1)^{F_R},
\label{rel sigma2}
\end{equation}
where $F_R$ denotes the `space-time fermion number' (mod $2$) 
from the right-mover. 
The operator $(-1)^{F_R}$ acts as the sign flip on all the states 
belonging to the right-moving R-sector.

~

As long as the $M^4$ part or $\psi^{2,3}$ are kept intact,
other possibilities essentially reduce to one of these two. 
%
%
%
%
The chiral blocks of the right-moving fermions 
in the eight-dimensional transverse part
are then 
determined in the same way as in the  bosonic $T^4$ sector:
we first evaluate the trace over the untwisted sector with the insertion of $\sigma^b$,  
and next require the modular covariance. 

For case {(i)}, we then have
the desired chiral blocks $\overline{f_{(a,b)}(\tau)}$ with 
\begin{eqnarray}
 f_{(a,b)}(\tau)
& =  &
\left\{
\begin{array}{ll}
(-1)^{\frac{a}{2}}
\left\{
\left(\frac{\th_3}{\eta}\right)^2\left(\frac{\th_4}{\eta}\right)^2
- \left(\frac{\th_4}{\eta}\right)^2\left(\frac{\th_3}{\eta}\right)^2
+0 \right\}
 &  ~~ (a\in 2\bz,~ b\in 2\bz+1) ,\\
(-1)^{\frac{b}{2}}
\left\{
\left(\frac{\th_3}{\eta}\right)^2\left(\frac{\th_2}{\eta}\right)^2 +0 
- \left(\frac{\th_2}{\eta}\right)^2\left(\frac{\th_3}{\eta}\right)^2
\right\}
 &  ~~ (a\in 2\bz+1,~ b\in 2\bz) ,\\
-  e^{\frac{i\pi}{2}ab}
\left\{ 0+
\left(\frac{\th_2}{\eta}\right)^2\left(\frac{\th_4}{\eta}\right)^2
- \left(\frac{\th_4}{\eta}\right)^2\left(\frac{\th_2}{\eta}\right)^2
\right\}
 &  ~~ (a\in 2\bz+1,~ b\in 2\bz+1) , \\
\left(\frac{\th_3}{\eta}\right)^4
- \left(\frac{\th_4}{\eta}\right)^4
-\left(\frac{\th_2}{\eta}\right)^4
 & ~~ (a \in 2\bz, ~ b\in 2\bz) .
\end{array}
\right.
\label{fab}
\end{eqnarray}
Each term from the left to the right corresponds to the NS, $\tNS$, 
and R sector, respectively,
where the `$\tNS$' denotes the NS-sector with $(-1)^f$ inserted 
($f$ is the world-sheet fermion number).
These trivially vanish as expected from  the space-time SUSY. 

We note that in the fermionic sectors
the modular covariance means\footnote
  {Since $f_{(a,b)} (\tau)$ vanish, \eqn{mc fab} may appear to be subtle. 
Hence, we present a more rigid interpretation of modular covariance in Appendix B.  
}:
\begin{eqnarray}
 && f_{a,b}(\tau)|_S 
= f_{(b,-a)}(\tau), 
\hspace{1cm}
f_{(a,b)}(\tau)|_T 
= - e^{-2\pi i \frac{1}{6}} f_{(a,a+b)}(\tau) ,
\label{mc fab}
\end{eqnarray}
with the phase for the T-transformation.
Since the total blocks for the transverse fermions consist of 
$\overline{f_{(a,b)}(\tau)}$ and the left-moving part,
\begin{equation}
\cJ(\tau) \equiv \left(\frac{\th_3}{\eta}\right)^4
- \left(\frac{\th_4}{\eta}\right)^4 - \left(\frac{\th_2}{\eta}\right)^4, 
\label{Jacobi}
\end{equation} 
(\ref{mc fab}) indeed assures the proper modular covariance:
\begin{eqnarray}
 \left. \left[ \cJ(\tau) \overline{f_{(a,b)}(\tau)} \right] \right|_S 
= \cJ(\tau)\overline{f_{(b,-a)}(\tau)}, 
\hspace{1cm}
\left.
\left[ \cJ(\tau)
\overline{f_{(a,b)}(\tau)} \right] \right|_T 
= \cJ(\tau) \overline{f_{(a,a+b)}(\tau)} .
\label{mc fab 2}
\end{eqnarray}


We next consider the chiral blocks for case {(ii)}, which we denote  by 
$\overline{\f_{(a,b)}(\tau)}$.
In this case, the treatment of the R-sector needs a little more care.
First, from \eqn{sigma H'} we find that
\begin{equation}
\f_{(0,b)}(\tau) = f_{(0,b)}(\tau), ~~~ (\any b \in 2\bz+1) ,
\label{bf 0b}
\end{equation}
which are vanishing. 
On the other hand, the blocks for the sectors of $a,b \in 2\bz$ 
are non-trivially modified due to \eqn{rel sigma2}.
Again it is easy to  evaluate the trace over the $(0,b)$-sector, 
and by requiring 
the modular covariance (in the sense of \eqn{mc fab} 
or \eqn{mc fab refined}), we finally obtain
\begin{eqnarray}
 && \f_{(a,b)}(\tau) = f_{(a,b)}(\tau), ~~~ 
(a\in 2\bz+1~  \mbox{or}~
 b \in 2\bz+1), 
 \label{fab 2}
\end{eqnarray}
and 
\begin{equation}
 \f_{(a,b)}(\tau) = \left\{
\begin{array}{ll}
\left(\frac{\th_3}{\eta}\right)^4
- \left(\frac{\th_4}{\eta}\right)^4
- \left(\frac{\th_2}{\eta}\right)^4
& ~~ (a\in 4\bz, ~ b \in 4\bz) , \\
\left(\frac{\th_3}{\eta}\right)^4
- \left(\frac{\th_4}{\eta}\right)^4
+ \left(\frac{\th_2}{\eta}\right)^4
& ~~ (a\in 4\bz, ~ b \in 4\bz+2) , \\
\left(\frac{\th_3}{\eta}\right)^4
+ \left(\frac{\th_4}{\eta}\right)^4
- \left(\frac{\th_2}{\eta}\right)^4
& ~~ (a\in 4\bz+2, ~ b \in 4\bz) , \\
-\left\{\left(\frac{\th_3}{\eta}\right)^4
+ \left(\frac{\th_4}{\eta}\right)^4
+ \left(\frac{\th_2}{\eta}\right)^4 \right\}
& ~~ (a\in 4\bz+2, ~ b \in 4\bz+2). \\
\end{array}
\right.
\label{fab 2-2} 
\end{equation}
In contrast to $f_{(a,b)}$, these $\f_{(a,b)}$ are in general non-vanishing, which 
signals the SUSY breaking in the right-moving sector.
%
This completes our construction of the chiral building blocks.
These are used in the following sections.

~


\section{String Vacua on T-folds}
\label{StVac}

Now we construct type II string vacua 
by combining the building blocks derived in the previous section.
They are interpretable as describing the compactification on T-folds.

First, to describe the `base sector' for $S^1_{\rm base}$, 
we introduce  
the following  notation,
\begin{eqnarray}
 Z_{R, (w,m)}(\tau,\bar{\tau}) &: =& 
\frac{R}{\sqrt{\tau_2}|\eta(\tau)|^2} 
e^{-\frac{\pi R^2}{\tau_2}|w\tau+m|^2}, ~~~ (w,m \in \bz),
\label{Z wm}
\end{eqnarray}
where $R$ is the radius of the compactification and the integers $w$, $m$ 
are identified as the spatial and temporal winding numbers. 
In terms of these, we find%
\footnote
   {Here we adopt the conventional normalization 
of the trace for  the  CFT for $\br_{\msc{base}}$,
$$
\tr_{\msc{base}} \left[q^{L_0-\frac{c}{24}} \overline{q^{\tL_0-\frac{c}{24}}}\, \right]
=\frac{R}{\sqrt{\tau_2} \left|\eta\right|^2}.
$$
This means that we start with $S^1_{NR}$ for the base CFT
with an arbitrary integer $N$, and 
regard the insertion of the shift operators $(\cT_{2\pi R})^m$ as 
 implementing 
the $\bz_N$-orbifolding. } 
\begin{equation}
\tr_{\msc{base}} \left[\left(\cT_{2\pi R}\right)^m \, q^{L_0-\frac{c}{24}} 
\overline{q^{\tL_0-\frac{c}{24}}}\, \right]
= Z_{R, (0,m)}(\tau,\bar{\tau}),
\label{cT^m insertion}
\end{equation}
and the torus partition function of a free compact boson with radius
$R$ reads
\begin{eqnarray}
 Z_{R} (\tau,\bar{\tau}) &=& \sum_{w,m\in \bsz} 
Z_{R,(w,m)}(\tau,\bar{\tau}).
\label{Z R}
\end{eqnarray} 

To calculate the total partition function, we proceed as follows:
First, we evaluate 
\begin{eqnarray}
Z_{(0,m)}(\tau,\bar{\tau}) &\equiv& \tr_{w=0~\msc{sector}} 
\left[\left( \cT_{2\pi R} \otimes \sigma\right)^m \, q^{L_0-\frac{c}{24}} 
\overline{q^{\tL_0-\frac{c}{24}}}\, \right]
\nn
& = & Z_{R,(0,m)}(\tau,\bar{\tau}) \, 
\tr_{\msc{untwisted}} \left[\sigma^m \, q^{L_0-\frac{c}{24}} 
\overline{q^{\tL_0-\frac{c}{24}}}\, \right].
\label{Z 0m total}
\end{eqnarray}
%
Second, we
extend \eqn{Z 0m total} to the partition function of the general 
winding sector $Z_{(w,m)}(\tau,\bar{\tau})$ 
by requiring the modular covariance. 
It is straightforward to perform this, given the relevant building blocks  
in the previous section.
These two steps are  also in parallel with the previous section.
Finally, we obtain the total partition function
by summing over the winding numbers 
$w,m \in \bz$ along the base circle 
as 
\begin{equation}
Z(\tau,\bar{\tau}) = \sum_{w,m\in \bz}\, Z_{(w,m)}(\tau,\bar{\tau}).
\end{equation}
%




\subsection{Supersymmetric Vacua}

In this way, we can construct string vacua, depending on the combination 
of the bosonic $T^4$ sector (1-3) in section \ref{bosonic sector}
and the transverse fermionic sector (i, ii) in section \ref{fermionic sector}.
All these are supersymmetric.

As the first example,  we consider 
$T^4[SO(8)]$ in the background \eqn{setup}.
Choosing case {(i)} for the fermionic sector, we obtain 
the torus partition function as 
\begin{eqnarray}
\hspace{-5mm}
Z(\tau, \bar{\tau}) &=& \frac{1}{4} Z^{\msc{tr}}_{M^4\times S^1}(\tau,\bar{\tau}) 
\sum_{w,m \in \bz}\,  Z_{R,(w,m)}(\tau,\bar{\tau}) \,
F^{T^4[SO(8)]}_{(w,m)}(\tau, \bar{\tau}) \,  \cJ (\tau)\,
 \overline{f_{(w,m)}(\tau)},
\label{SUSY T-fold 1}
\end{eqnarray}
where $Z^{\msc{tr}}_{M^4 \times S^1}(\tau,\bar{\tau})$ 
denotes
the bosonic partition function for the transverse part of $M^4 \times S^1$-sector. 
$\cJ(\tau)$ is the contribution from the left-moving free fermions defined in \eqn{Jacobi},
and the overall factor $1/4$ is due to the chiral GSO projections. 
This is manifestly modular invariant by construction and defines a superstring vacuum, 
which 
preserves $3/4$ of the space-time SUSY, that is, 
16 supercharges from the left-mover and 8 supercharges from the right-mover. 

For case {(ii)}, 
we replace $f_{(w,m)}(\tau)$ in \eqn{SUSY T-fold 1} with 
$\f_{(w,m)}(\tau)$ given in \eqn{fab 2}, \eqn{fab 2-2},
and obtain the torus partition function 
\begin{eqnarray}
\hspace{-5mm}
Z(\tau, \bar{\tau}) &=& \frac{1}{4} Z^{\msc{tr}}_{M^4 \times S^1}(\tau,\bar{\tau}) 
\sum_{w,m  \in \bz}\,  Z_{R,(w,m)}(\tau,\bar{\tau})  \,F^{T^4[SO(8)]}_{(w,m)}(\tau, \bar{\tau}) 
\, \cJ (\tau) \, 
 \overline{\f_{(w,m)}(\tau)}.
\label{SUSY T-fold 2}
\end{eqnarray}
This time, we are left with 
the 1/2 space-time SUSY that originates 
only from the left-mover.\footnote
   {See the discussions given in section \ref{non-SUSY vacua} 
   for the counting of unbroken supercharges in more detail. }

It is straightforward to construct the string vacua 
in other four cases based on $T^2[SO(4)] \times T^2[SO(4)]$:
one has only to replace the bosonic building blocks 
$F^{T^4[SO(8)]}_{(w,m)}(\tau, \bar{\tau}) $
in the  above with 
\eqn{Fab T2 T2} or \eqn{tFab T2 T2}
without any changes in other sectors.


\subsection{Non-SUSY String Vacua with Vanishing Cosmological Constant}
\label{non-SUSY vacua}

An interesting modification of the half SUSY vacuum 
represented by \eqn{SUSY T-fold 2} is to replace 
the base circle along the $X^5$-direction 
with the Scherk-Schwarz one \cite{SS1,SS2}. This means that we 
implement 
the orbifolding of the background \eqn{setup} by the twist operator\footnote
  {If following the notion of the original Scherk-Schwarz compactification, 
it would be better to introduce 
$$
g' : = \cT_{2\pi R} \otimes (-1)^{F^S} \otimes \sigma 
\equiv \cT_{2\pi R} \otimes (-1)^{F_L } \otimes 
\left[
\sigma (-1)^{F_R} \right],
$$
instead of \eqn{twist g}, 
where $F^S \equiv F_L + F_R$ is the space-time fermion number.
However, the argument given here is almost unchanged even in that case,
and especially, we end up with 
the same torus partition function \eqn{non-SUSY vacuum}. 
}
\begin{equation}
g := \cT_{2\pi R} \otimes (-1)^{F_L} \otimes \sigma ,
\label{twist g}
\end{equation}
where $(-1)^{F_L}$ acts as the sign flip on any states 
of the left-moving Ramond sector. 
Again $\sigma$ denotes the chiral reflection for the $T^4$-sector and 
is assumed to satisfy $\sigma^2 = (-1)^{F_R}$ as for \eqn{SUSY T-fold 2}.
The action of the twist operators $g^n$ is summarized in Table \ref{Table1}.


\begin{table}[t]
\begin{center}
\begin{tabular}{|c|c|c|c|c|}
\hline
      & base $(X^5)$ & $T^4$ ($X^{6,7,8,9})$ & left-moving fermions 
      & right-moving fermions    \\ \hline\hline
$g^{4n}$   & $\cT_{2\pi (4n)R}$  & $\bm{1}$    & $ \bm{1}$  &  $\bm{1}$  \\ \hline
$g^{4n+1}$   & $\cT_{2\pi (4n+1)R}$  & $\sigma $    & $(-1)^{F_L}$   &  $\sigma $ \\ \hline
$g^{4n+2}$   & $\cT_{2\pi (4n+2)R}$  & $ \bm{1}$    & $\bm{1}$   &   $ (-1)^{F_R}$ \\ \hline
$g^{4n+3}$   & $\cT_{2\pi (4n+3)R}$  & $\sigma $    &  $(-1)^{F_L}$ 
 &   $(-1)^{F_R} \sigma $ \\ \hline
\end{tabular}
\caption{Action of the twist operators $g^n$}
\label{Table1}
\end{center}
\end{table} 

This modification 
leads to the following torus partition function,
\begin{equation}
Z(\tau,\bar{\tau}) = \frac{1}{4} Z^{\msc{tr}}_{M^4\times S^1}(\tau,\bar{\tau})  \,
\sum_{w,m \in \bz}\, Z_{R, (w,m)}(\tau,\bar{\tau}) \, F^{T^4[SO(8)]}_{(w,m)}(\tau, \bar{\tau}) 
\, \f_{(2w,2m)}(\tau)\, 
 \overline{\f_{(w,m)}(\tau)}.
\label{non-SUSY vacuum}
\end{equation}
Here, the chiral blocks for left-moving fermions have been replaced with $\f_{(2w,2m)}(\tau)$ as in \eqn{SUSY T-fold 2} due to 
the extra twisting $(-1)^{F_L}$.
One can confirm that this partition  function 
vanishes for each winding sector, similarly to usual supersymmetric string vacua.
Indeed, 
$\overline{\f_{(w,m)}(\tau)}=0$ for $\any w \in 2\bz+1$ or $\any m \in 2\bz+1$,
while $\f_{(2w,2m)}(\tau)=0$ for $\any w, m \in 2\bz$.
Then, we see a bose-fermi cancellation at each mass level of the string spectrum, 
after making the Poisson resummation with respect to the temporal winding $m$ in a standard fashion. 
We will observe this aspect explicitly in section 4.
Thus, the vacuum energy or the cosmological constant in  space-time  vanishes 
at the one-loop level.

A remarkable fact here is that the space-time SUSY is nonetheless completely broken:
\begin{itemize}
\item 

For $w=0$, 
only the supercharges  
commuting with the orbifold projection 
$\frac{1}{4}\sum_{n\in \bz_4} \, \left. g^n \right|_{\msc{fermion}}$
would be preserved. 
However, since the relevant projection includes both $(-1)^{F_L}$ and $(-1)^{F_R}$, 
all the supercharges in the unorbifolded theory cannot commute with it. This implies that 
all the supercharges 
from this sector are projected out. 
\item 
For $w\neq 0$, 
if we had a supercharge, 
we would observe a bose-fermi cancellation between two sectors with 
winding numbers $w'$ and $w'+w$ for $\any w' \in \bz$,  which would imply 
\begin{equation}
Z^{(\sNS,\sNS)}_{w'}(\tau,\bar{\tau}) + Z^{(\sR,\sR)}_{w'}(\tau, \bar{\tau}) = 
- \left\{ Z^{(\sNS,\sR)}_{w'+w}(\tau,\bar{\tau}) 
+ Z^{(\sR,\sNS)}_{w'+w}(\tau,\bar{\tau}) \right\}.
\label{bose-fermi w}
\end{equation}
However, 
we explicitly confirm, as expected, in section \ref{unitarity}
that such relations never hold for the partition function \eqn{non-SUSY vacuum}  
due to the factor $Z_{R, (w,m)}(\tau,\bar{\tau})$ from the base circle. 
\end{itemize}

~


Here, it would be worthwhile to emphasize a crucial role 
of the shift operator  $\left. \cT_{2\pi R} \right|_{\msc{base}}$ 
 in the above argument. Obviously, one has 
a vanishing partition function even without 
$Z_{R, (w,m)}(\tau,\bar{\tau})$:
\begin{equation}
\tilde{Z}(\tau,\bar{\tau}) = \frac{1}{4\cdot 4}  
Z^{\msc{tr}}_{M^4\times S^1\times S^1} (\tau,\bar{\tau}) \,
\sum_{a,b \in \bz_4}\,  F^{T^4[SO(8)]}_{(a,b)}(\tau, \bar{\tau}) 
\, \f_{(2a,2b)}(\tau)\, 
 \overline{\f_{(a,b)}(\tau)}.
\label{failed model}
\end{equation}
For the untwisted sector with $a=0$, all the supercharges are 
projected out in the same way as above.
However, new Ramond vacua can appear from 
the $a\neq 0$ sectors in this case,%
\footnote
   {In fact, the orbifolding by $(-1)^{F_L}$ (or $(-1)^{F_R}$) acts 
   as the `chirality flip' of the Ramond sector,
which transfers the type IIA (IIB) vacua to the type IIB (IIA) ones 
similarly to T-duality. See {\em e.g.} \cite{GabS}.
}
and the space-time SUSY revives eventually. 
The inclusion of  $\left. \cT_{2\pi R} \right|_{\msc{base}}$ 
was a very simple way to exclude such a possibility, 
since supercharges cannot carry winding charges generically, as pointed out above.
This is also in accord with an intuition that  in the twisted sectors masses are lifted up
by the winding charges. 



\subsection{Asymmetric/Generalized Orbifolds and T-folds}

We have explicitly constructed the non-geometric superstring 
vacua/partition functions,
\eqn{SUSY T-fold 1}, \eqn{SUSY T-fold 2}, \eqn{non-SUSY vacuum}
for the asymmetric orbifolds associated with the chiral reflection. 
In this subsection, we would like to comment on 
the relation to the construction of T-folds in
\cite{HW,Tan:2015nja}.  In these works,  
the T-duality twists are accompanied by extra phases, 
so that the full operator product expansion (OPE), 
not only the chiral one, respects
the invariance under the twist:
supposed that two vertex operators including both the left and right movers
are invariant, their OPE yields invariant operators.
This is in accord with the ordinary principle of orbifolding by symmetries.
The construction of  (\ref{tFab T2 T2}) includes such phases 
and the resultant models represent the T-folds in this sense.
Asymmetric orbifolding, however, generally respects the chiral OPE only, 
and belongs to a different class.

Here, we recall  
that, from the CFT point of view, T-duality is in general 
an isomorphism between different Hilbert spaces, which keeps 
the form of the Hamiltonian invariant. 
At the self-dual point, it acts within a single Hilbert space,
but is not yet an ordinary symmetry, since the transformation 
to the dual fields is non-local.
Thus, it may not be obvious if the OPE should fully respect the invariance 
under the T-duality twists.
Indeed, in the case of the critical Ising model, the OPE of the order and  
disorder fields, which are non-local to each other,  reads 
\begin{equation}
\sigma(z,\bar{z}) \mu(0,0) \sim |z|^{-1/4} 
\bigl[ \omega z^{1/2} \psi(0) + \bar\omega \bar{z}^{1/2} \bar\psi(0)\bigr],
\end{equation}
where 
$\psi, \bar\psi$ are the free fermions, 
$\omega = \frac{1}{\sqrt{2}}e^{\frac{i\pi}{4} }$
and $\bar\omega$ is its complex conjugate.
Under the Kramers-Wannier duality (T-duality), these fields are mapped as 
$(\sigma, \mu, \psi, \bar\psi) \to (\mu, \sigma, \psi, -\bar\psi)$.
One then finds that
the OPE of two invariant fields $ (\sigma+\mu)(z,\bar{z}) (\sigma+\mu)(0,0)$
yields non-invariant fields, since the diagonal part  $\sigma \sigma + \mu \mu$ 
yields invariant ones.

In addition, we  note that 
sensible CFTs may be obtained from the twists 
by transformations which are not the full symmetries.
We refer to such CFTs as ``generalized orbifold" CFTs, according
to \cite{Frohlich:2009gb,Carqueville:2012dk,Brunner:2013ota}
where such CFTs 
are studied in the context of the topological 
conformal interfaces 
\cite{Wong:1994np,Petkova:2000ip,Bachas:2001vj,Bachas:2004sy}. 
An application to non-geometric backgrounds has been discussed 
in \cite{SatohS}.
Even though the twists are not necessarily by
the full symmetries, the transformations 
may need to commute with the Hamiltonian, since 
the position of the twist operators matters otherwise.
In this terminology, general asymmetric orbifold models and hence 
ours based on the twists without the extra phases belong to this class.
In any case, our resultant models are consistent in that they are 
modular invariant
and, as shown in the next section, have sensible spectra.
 
Taking these into account, we expect that the world-sheet CFTs for T-folds 
are generally given by the asymmetric/generalized orbifold CFTs, and that 
our asymmetric orbifolds without, as well as with, 
the extra phases also represent T-folds, as we have assumed so far
(see also
\cite{Dabholkar:2002sy,Hellerman:2002ax,Flournoy:2004vn,Dabholkar:2005ve,
Flournoy:2005xe,Condeescu:2012sp,Condeescu:2013yma}). 
It would be an interesting issue if 
all these non-geometric models have the corresponding supergravity description 
as low-energy 
effective theory of T-folds. As is discussed shortly, 
the difference of the spectra due to the phases typically appear 
in the massive sector. However, 
the massless spectra can also differ, for example, at special points 
of the moduli, and thus supergravity may distinguish them.

Regarding the interpretation as T-folds, we also note that
the chiral reflections
both for $T^4[SO(8)] $ and $ T^2[SO(4)] \times T^2[SO(4)] $ 
are indeed realized as self-dual 
 $O(4,4,{\mathbb Z})$ transformations which leave 
background geometries invariant.
The elements of $ O(4,4,{\mathbb Z})$ act as  ${\mathbb Z}_2$ transformations 
in the untwisted bosonic 
sector as expected, whereas they do not  
generally in other sectors, for example, 
in the fermionic sectors (see also \cite{HW,Aoki:2004sm}). This, however, 
is not a contradiction:
that means that such sectors are in different representations.

~


\section{Analysis on Spectra}

\subsection{Massless Spectra in the Untwisted Sectors}

To clarify the physical content of 
the non-SUSY vacuum with the bose-fermi cancellation 
\eqn{non-SUSY vacuum}, let us  examine the massless spectrum 
in the untwisted sector $(w=0)$ that survives in the low energy physics. 
The massless states from the twisted sectors $(w \neq 0)$
 can appear 
only at the special radius $R$ (see subsection \ref{tachyon}).

We first note the fact that
all the right-moving Ramond vacua are projected out by the orbifold action $g$;
recall $\sigma^2 = (-1)^{F_R}$ for the world-sheet fermions.
Therefore, the candidates of the bosonic and fermionic massless states 
only reside in  the $(\NS, \NS)$ and $(\R, \NS)$-sectors, respectively.
It is thus enough to search the $(\NS, \NS)$ and $(\R, \NS)$ massless states 
invariant under the action of 
$ (-1)^{F_L} \otimes \sigma$ 
within the Hilbert space of the unorbifolded  theory. 
In this way, one can easily write down the massless spectrum.
We exhibit it in  Table \ref{massless spectrum non-SUSY}.%
\footnote
   {Here, the `14 (pseudo) scalars' include the dilaton and 
   the 4-dimensional 
   axion field (dual of $B_{\mu \nu}$), which universally exist. 
}
Since our background includes the $S^1$-factor ($X^4$-direction) 
that is kept intact under the orbifolding, 
it is evident by considering T-duality
 that the type IIA and type IIB vacua lead us to the same massless spectra in 
 4 dimensions. Thus, we do not 
specify here which we are working on. 


\begin{table}[t] 
\begin{center}
	
	\begin{tabular}{|c|ccc|c|}
\hline
 spin structure & left & & right &  4D fields   \\
 \hline\hline
(NS, NS) & $\psi ^\mu _{-1/2}\ket 0 $& $\otimes $ & 
$ \tilde \psi ^\nu  _{-1/2}\ket 0 $
&  
graviton, 
8 vectors,  \\
&
$(\mu =2,...,9)$ & & $(\nu =2,..., 5) $&  
14 (pseudo) scalars 
\\
\hline
(R, NS) &
$\ket{\ep_0, \ep_1, \ep_2, \ep_3}$& $\otimes $ & 
$\tilde \psi ^\nu  _{-1/2}\ket 0  $
 & 16 Weyl fermions  \\
&  & & $(\nu =6,...,9)$ &   \\
\hline
\end{tabular}
\caption{Massless spectrum of the non-SUSY vacuum 
 \eqn{non-SUSY vacuum}. 
} 
\label{massless spectrum non-SUSY} 
\end{center}
\end{table}
It is evident from Table 2 that we have the same number of the
massless bosonic and fermionic degrees of freedom. 
Nevertheless, there are no 4-dimensional  gravitini, 
reflecting the absence of the space-time SUSY. 


For comparison, it would be useful to 
exhibit the massless spectra 
in the untwisted sector for the 3/4-SUSY vacuum \eqn{SUSY T-fold 1}
and the 1/2-SUSY vacuum \eqn{SUSY T-fold 2}.
We present them in 
Table \ref{massless spectrum SUSY T-fold 1} and 
Table \ref{massless spectrum SUSY T-fold 2}.
\begin{table}[t]
\begin{center}
	
	\begin{tabular}{|c|ccc|c|}
\hline
 spin structure & left & & right &  4D fields   \\
 \hline\hline
(NS, NS) & $\psi ^\mu _{-1/2}\ket 0 $& $\otimes $ & 
$ \tilde \psi ^\nu  _{-1/2}\ket 0 $
&  
graviton, 
8 vectors,  \\
&
$(\mu =2,...,9)$ & & $(\nu =2,..., 5) $&  
14 (pseudo) scalars 
\\
\hline
(R, R) & $\ket{\ep_0, \ep_1, \ep_2, \ep_3}$& $\otimes $ & 
$\ket{\tep_0, \tep_1, \tep_2, \tep_3}$ &
8 vectors,
\\
 & & & $(\tep_2 + \tep_3=0)$ &
 16 (pseudo) scalars
 \\
\hline
(R, NS) &
$\ket{\ep_0, \ep_1, \ep_2, \ep_3}$& $\otimes $ & 
$\tilde \psi ^\nu  _{-1/2}\ket 0  $
 & 8 gravitini,   \\
&  & & $(\nu =2,...,5)$ &  8 Weyl fermions \\
\hline
(NS, R) &
$\psi ^\mu _{-1/2}\ket 0 $ & $\otimes $ & 
$\ket{\tep_0, \tep_1, \tep_2, \tep_3}$
 & 4 gravitini,   \\
& $(\mu =2,...,9)$ & & $(\tep_2 + \tep_3=0)$ &  12 Weyl fermions \\
\hline
\end{tabular}
\caption{Massless spectrum of the SUSY vacuum \eqn{SUSY T-fold 1} } 
\label{massless spectrum SUSY T-fold 1} 
\end{center}
\end{table}


\begin{table}[htbp]
\begin{center}
	
	\begin{tabular}{|c|ccc|c|}
\hline
 spin structure & left & & right &  4D fields   \\
 \hline\hline
(NS, NS) & $\psi ^\mu _{-1/2}\ket 0 $& $\otimes $ & 
$ \tilde \psi ^\mu  _{-1/2}\ket 0 $
&  
graviton, 
8 vectors,  \\
&
$(\mu =2,...,9)$ & & $(\mu =2,..., 5) $&  
14 (pseudo) scalars 
\\
\hline
(R, NS) &
$\ket{\ep_0, \ep_1, \ep_2, \ep_3}$& $\otimes $ & 
$\tilde \psi ^\mu  _{-1/2}\ket 0  $
 & 8 gravitini,   \\
&  & & $(\mu =2,...,5)$ &  8 Weyl fermions \\
\hline
\end{tabular}
\caption{Massless spectrum of the SUSY vacuum \eqn{SUSY T-fold 2} } 
\label{massless spectrum SUSY T-fold 2} 
\end{center}
\end{table}



\subsection{Unitarity}
\label{unitarity}

The torus partition functions we constructed in the previous section 
include the non-trivial phase factors which originate from the requirement of the modular covariance
and depend on the winding numbers along the base circle.
Thus, it may  not be so obvious whether the spectrum is  unitary in each vacuum,
though that is evident in the untwisted sector with $w=0$ by construction.

An explicit way to check the unitarity is to examine the string spectrum by 
the Poisson resummation of the relevant partition function 
with respect to the temporal winding 
$m$ along the base circle. 
To this end, we decompose the partition functions 
with respect to the spatial winding $w$ and the spin 
structures, and factor out the component of 
$Z^{\msc{tr}}_{M^4\times S^1}$:
\begin{equation}
Z(\tau,\bar{\tau}) = \frac{1}{4} Z^{\msc{tr}}_{M^4\times S^1}(\tau,\bar{\tau})  \,
\sum_{s, \ts}\, \sum_{w\in \bz} \, Z^{(s, \ts)}_w (\tau, \bar{\tau}),
\label{decomp Z}
\end{equation}
where $s, \ts = \NS, \, \R$ denote the left and right-moving 
spin structures.


For instance, let us pick up the non-SUSY vacuum built from $T^4[SO(8)]$
given by \eqn{non-SUSY vacuum}.
Making the Poisson resummation, 
we find that each function  $Z^{(s, \ts)}_w (\tau, \bar{\tau})$
with fixed $w$ becomes
as follows:
\begin{itemize}
\item $w \in 4\bz$ ;
\begin{eqnarray}
&& 
 Z^{(\sNS, \sNS)}_w (\tau, \bar{\tau})  =
- Z^{(\sR, \sNS)}_w (\tau, \bar{\tau}) 
= \frac{1}{4}  \sum_{n\in \bz} \,
\overline{q^{\frac{1}{4} \left(\frac{n}{2R} - R w\right)^2}} q^{\frac{1}{4}
 \left(\frac{n}{2R} + R w\right)^2} 
\nn
&&
\hspace{1cm}  
\times \left\{ 
\left|\frac{\th_3}{\eta}\right|^8  + \left|\frac{\th_4}{\eta}\right|^8
+\left|\frac{\th_2}{\eta}\right|^8
\right\}
\left|\left(\frac{\th_3}{\eta} \right)^4 - \left(\frac{\th_4}{\eta} \right)^4  \right|^2,
\label{Z NSNS 4Z}
\\
&& 
 Z^{(\sR, \sR)}_w (\tau, \bar{\tau}) 
= - Z^{(\sNS, \sR)}_w (\tau, \bar{\tau}) 
= \frac{1}{4}  \sum_{n\in \bz+\frac{1}{2}}\, 
\overline{q^{\frac{1}{4} \left(\frac{n}{2R} - R w\right)^2}} q^{\frac{1}{4} 
\left(\frac{n}{2R} + R w\right)^2} 
\nn
&&
\hspace{1cm}  
\times 
\left\{ 
\left|\frac{\th_3}{\eta}\right|^8  + \left|\frac{\th_4}{\eta}\right|^8
+\left|\frac{\th_2}{\eta}\right|^8
\right\}
\left|\frac{\th_2}{\eta} \right|^8 .
\label{Z RR 4Z}
\end{eqnarray}
%
%
%
%
\item $w \in 4\bz+2 $ ;
\begin{eqnarray}
&& 
 Z^{(\sNS, \sNS)}_w (\tau, \bar{\tau}) =
- Z^{(\sR, \sNS)}_w (\tau, \bar{\tau})
=
\frac{1}{4}  \sum_{n\in \bz+ \frac{1}{2}} 
\overline{q^{\frac{1}{4} \left(\frac{n}{2R} - R w\right)^2}} q^{\frac{1}{4} 
\left(\frac{n}{2R} + R w\right)^2} 
\nn
&&
\hspace{5mm}  
\times \left\{ 
\left|\frac{\th_3}{\eta}\right|^8  + \left|\frac{\th_4}{\eta}\right|^8
+\left|\frac{\th_2}{\eta}\right|^8
\right\}
\overline{
\left\{ \left(\frac{\th_3}{\eta} \right)^4 + \left(\frac{\th_4}{\eta} \right)^4\right\} 
}
\left\{ \left(\frac{\th_3}{\eta} \right)^4 - \left(\frac{\th_4}{\eta} \right)^4\right\} ,
\label{Z NSNS 4Z+2}
\\
&& 
 Z^{(\sR, \sR)}_w (\tau, \bar{\tau}) 
= - Z^{(\sNS, \sR)}_w (\tau,\bar{\tau})
= \frac{1}{4}  \sum_{n\in \bz}\, 
\overline{q^{\frac{1}{4} \left(\frac{n}{2R} - R w\right)^2}} q^{\frac{1}{4}
 \left(\frac{n}{2R} + R w\right)^2} 
\nn
&&
\hspace{1cm}  
\times 
\left\{ 
\left|\frac{\th_3}{\eta}\right|^8  + \left|\frac{\th_4}{\eta}\right|^8
+\left|\frac{\th_2}{\eta}\right|^8
\right\}
\left|\frac{\th_2}{\eta} \right|^8.
\label{Z RR 4Z+2}
\end{eqnarray}
%
%
\item $w\in 2\bz+1$ ;
\begin{eqnarray}
&& 
Z^{(\sNS, \sNS)}_w (\tau, \bar{\tau}) =
- Z^{(\sNS, \sR)}_w (\tau, \bar{\tau}) 
=
\frac{1}{4} \sum_{r\in \bz_2}\, \sum_{n\in \bz} 
\overline{q^{\frac{1}{4} \left(\frac{n}{2R} - R w\right)^2}} q^{\frac{1}{4} 
\left(\frac{n}{2R} + R w\right)^2} 
\nn
&&
\hspace{5mm}  \times (-1)^{rn}
\overline{\left(\frac{\th_2 \th_3(\frac{r}{2})}{\eta^2} \right)^4} 
\left\{ (-1)^r
\left(\frac{\th_3(\frac{r}{2})}{\eta}\right)^4 
+ \left(\frac{\th_2}{\eta}\right)^4 \right\}
\left\{\left(\frac{\th_3}{\eta}\right)^4  + \left(\frac{\th_4}{\eta}\right)^4\right\},
\label{Z NSNS 2Z+1}
\\
&& 
 Z^{(\sR, \sR)}_w (\tau, \bar{\tau}) = 
- Z^{(\sR, \sNS)}_w (\tau, \bar{\tau}) =
\frac{1}{4} \sum_{r\in \bz_2}\, \sum_{n\in \bz} 
\overline{q^{\frac{1}{4} \left(\frac{n}{2R} - R w\right)^2}} q^{\frac{1}{4}
 \left(\frac{n}{2R} + R w\right)^2} 
\nn
&&
\hspace{5mm}  \times (-1)^{rn}
\overline{\left(\frac{\th_2 \th_3(\frac{r}{2})}{\eta^2} \right)^4} 
\left\{ 
\left(\frac{\th_3(\frac{r}{2})}{\eta}\right)^4 
+ (-1)^r\left(\frac{\th_2}{\eta}\right)^4 \right\}
\left(\frac{\th_2}{\eta}\right)^4.
\label{Z RR 2Z+1}
\end{eqnarray}
Here, we denoted $\th_i \equiv \th_i(\tau,0)$, 
and $\th_3(\frac{r}{2}) \equiv \th_3(\tau, \frac{r}{2})$,
and made use of the identity $\th_3^4-\th_4^4- \th_2^4=0$.
\end{itemize}
%
As expected, all of these partition functions 
are suitably $q$-expanded
so as to be consistent with the unitarity. 

The SUSY T-fold vacua \eqn{SUSY T-fold 1}, \eqn{SUSY T-fold 2} 
are similarly analyzed. 
For the 3/4-SUSY vacuum \eqn{SUSY T-fold 1}, 
$Z^{(s, \ts)}_w (\tau, \bar{\tau})$  becomes as follows:
\begin{itemize}
\item $w\in 2\bz$ ;
\begin{equation}
Z^{(\sNS, \sNS)}_w (\tau, \bar{\tau}) = Z^{(\sR, \sR)}_w (\tau, \bar{\tau})
= - Z^{(\sR, \sNS)}_w (\tau, \bar{\tau}) 
= - Z^{(\sNS, \sR)}_w (\tau, \bar{\tau})   
= \eqn{Z NSNS 4Z} .
\label{SUSY 1 Z 2Z} 
\end{equation}
\item $w\in 2\bz+1$ ;
\begin{equation}
Z^{(\sNS, \sNS)}_w (\tau, \bar{\tau}) = Z^{(\sR, \sR)}_w (\tau, \bar{\tau})
= - Z^{(\sR, \sNS)}_w (\tau, \bar{\tau}) 
= - Z^{(\sNS, \sR)}_w (\tau, \bar{\tau})   
= \eqn{Z RR 2Z+1}.
\label{SUSY 1 Z 2Z+1}
\end{equation}
\end{itemize}
%
%
On the other hand, 
for the 1/2-SUSY vacuum \eqn{SUSY T-fold 2}, 
we find the following:
\begin{itemize}
\item $w\in 4\bz$ ;
\begin{eqnarray}
&& Z^{(\sNS, \sNS)}_w (\tau, \bar{\tau})  
= - Z^{(\sR, \sNS)}_w (\tau, \bar{\tau}) 
= \eqn{Z NSNS 4Z},
\nn
&& Z^{(\sR, \sR)}_w (\tau, \bar{\tau})
= - Z^{(\sNS, \sR)}_w (\tau, \bar{\tau})   
= \eqn{Z RR 4Z}.
\label{SUSY 2 Z 4Z}
\end{eqnarray}
%
\item $w\in 4\bz+2 $ ;
\begin{eqnarray}
&& Z^{(\sNS, \sNS)}_w (\tau, \bar{\tau})  
= - Z^{(\sR, \sNS)}_w (\tau, \bar{\tau}) 
= \eqn{Z NSNS 4Z+2},
\nn
&& Z^{(\sR, \sR)}_w (\tau, \bar{\tau})
= - Z^{(\sNS, \sR)}_w (\tau, \bar{\tau})   
= \eqn{Z RR 4Z+2} .
\label{SUSY 2 Z 4Z+2}
\end{eqnarray}
%
\item $w\in 2\bz+1$ ;
\begin{equation}
Z^{(\sNS, \sNS)}_w (\tau, \bar{\tau}) = Z^{(\sR, \sR)}_w (\tau, \bar{\tau})
= - Z^{(\sR, \sNS)}_w (\tau, \bar{\tau}) 
= - Z^{(\sNS, \sR)}_w (\tau, \bar{\tau})   
= \eqn{Z RR 2Z+1}.
\label{SUSY 2 Z 2Z+1}
\end{equation}
\end{itemize}
%
These analyses can be extended to other vacua  
built from $F_{(*,*)}^{T^2[SO(4)] \times T^2[SO(4)]}$ in 
\eqn{Fab T2 T2} or \\ 
$\tF_{(*,*)}^{T^2[SO(4)] \times T^2[SO(4)]}$  in 
\eqn{tFab T2 T2}. 
In each case, we obtain the unitary $q$-expansion
in a parallel way as above.

~

We remark that the above results 
\eqn{SUSY 1 Z 2Z} and \eqn{SUSY 1 Z 2Z+1} suggest that there are supercharges 
both from the left and right movers for the SUSY T-fold \eqn{SUSY T-fold 1}. 
Similarly,
\eqn{SUSY 2 Z 4Z}, \eqn{SUSY 2 Z 4Z+2} and \eqn{SUSY 2 Z 2Z+1} are 
consistent with the existence of the chiral SUSY
that originates only from the left-mover. 
Then, how about the non-SUSY vacuum \eqn{non-SUSY vacuum}?
We note that, for instance, 
\begin{eqnarray}
&&Z^{(\sNS, \sNS)}_w (\tau, \bar{\tau}) = - Z^{(\sR, \sNS)}_w (\tau, \bar{\tau}), 
\hspace{1cm} 
Z^{(\sR, \sR)}_w (\tau, \bar{\tau}) = - Z^{(\sNS, \sR)}_w (\tau, \bar{\tau}),
\nn
&& Z^{(\sNS, \sNS)}_w (\tau, \bar{\tau}) \neq - Z^{(\sNS, \sR)}_w (\tau, \bar{\tau}),
\hspace{1cm} 
Z^{(\sR, \sR)}_w (\tau, \bar{\tau}) \neq  - Z^{(\sR, \sNS)}_w (\tau, \bar{\tau}),
\label{rel non-SUSY 1}
\end{eqnarray}
for $w \in 2\bz$. 
These relations of the bose-fermi cancellation
look as if we had  
{\em left-moving\/} SUSY, in spite that 
no supercharges exist in the left-mover in fact.
On the other hand, we find 
\begin{eqnarray}
&& Z^{(\sNS, \sNS)}_w (\tau, \bar{\tau}) = - Z^{(\sNS, \sR)}_w (\tau, \bar{\tau}), 
\hspace{1cm} 
Z^{(\sR, \sR)}_w (\tau, \bar{\tau}) = - Z^{(\sR, \sNS)}_w (\tau, \bar{\tau}),
\nn
&& Z^{(\sNS, \sNS)}_w (\tau, \bar{\tau}) \neq - Z^{(\sR, \sNS)}_w (\tau, \bar{\tau}),
\hspace{1cm}
Z^{(\sR, \sR)}_w (\tau, \bar{\tau}) \neq - Z^{(\sNS, \sR)}_w (\tau, \bar{\tau}),
\label{rel non-SUSY 2}
\end{eqnarray}
for $w\in 2\bz+1$, 
which would appear to be consistent with 
{\em right-moving\/}  SUSY. 
We emphasize  that any supercharges  can never be compatible with   
both \eqn{rel non-SUSY 1} and \eqn{rel non-SUSY 2} at the same time.
It may be an interesting issue whether such a curious feature is common 
to the vacua showing  the bose-fermi cancellation without SUSY. 
%

We also point out that the bose-fermi cancellation in \eqn{bose-fermi w}
among {\em different \/} winding sectors does not happen (for arbitrary $w'$), 
as is clear 
from the explicit forms of the partition functions presented above.
Even at a special radius, the cancellation for arbitrary winding in \eqn{bose-fermi w}
is not possible.



\subsection{Absence of Winding Tachyons}
\label{tachyon}

Recall that our non-SUSY string vacuum \eqn{non-SUSY vacuum} 
from $T^4[SO(8)]$
has been constructed 
by including the Sherk-Schwarz type modification.
Therefore,  we would potentially face an issue of 
the instability caused by the winding tachyons
that are typical in the Sherk-Schwarz compactification. 
That would be implied by 
 the `wrong GSO projections' observed   in \eqn{Z NSNS 4Z+2}, 
 \eqn{Z NSNS 2Z+1}.%
\footnote
 {In the T-fold vacuum \eqn{SUSY T-fold 2}, despite the existence 
 of the space-time SUSY, 
we still find the wrong GSO fermions in the right-mover (with no SUSY), since 
$Z^{(\sNS,\sNS)}_w  \left( = - Z^{(\sR, \sNS)}_w\right) $ 
coincides with the partition function 
\eqn{Z NSNS 4Z+2} for $w \in 4\bz+2$.
Of course, one can confirm 
the absence of tachyonic modes in this model
by a similar argument given here.
}
However, the spectrum is in fact  free from the winding tachyons.

To show this, we first note that potentially dangerous states come from 
the winding sectors with $w\in 4\bz+2$ or $w\in 2\bz+1$, 
which are anticipated from the wrong GSO projections.
Among them, we further focus on the NS-NS sector, since 
the spectrum is lifted in the R-R sector due to the $\theta_2$-factors, 
and the partition functions in  the NS-R and R-NS  sectors
are the same as  for the NS-NS or the R-R sector up to sign.
From the partition functions \eqn{Z NSNS 4Z+2}, \eqn{Z NSNS 2Z+1}, 
we  then find the following: 
\begin{itemize}
\item 
For $w\in 4\bz+2$,  the wrong GSO states are in the right-mover.
The lightest excitations appear in the sectors of  $w=\pm 2$, 
the conformal weights of which read
\begin{equation}
h_L = \frac{1}{2} + 
      \frac{1}{4} \left(\frac{n}{2R}  \pm 2R \right)^2, \hspace{1cm}
h_R = \frac{1}{4} \left(\frac{n}{2R} \mp 2R \right)^2 ,
\end{equation}
with the KK momenta $n \in \bz+\frac{1}{2}$.
Their minima for the physical states are achieved by setting
$n= \mp \frac{1}{2}$, to give 
\begin{equation}
h_L = h_R = \frac{1}{2} 
+ \frac{1}{4} \left(\frac{1}{4R} -2R \right)^2 \geq \frac{1}{2}.
\label{lightest 1}
\end{equation}
This means that the winding states from these sectors 
are always massive except at 
the special radius $R= \frac{1}{2\sqrt{2}}$ of the base circle, where 
extra massless excitations appear.

\item For $w\in 2\bz+1$, 
the wrong GSO states are
in the left-mover. 
The lightest excitations appear in the sectors of $w=\pm 1$,
and the leading contribution from the $\theta$-part comes from 
$\th_3(\frac{r}{2}) = 1 + (-1)^r q^{\frac{1}{2}} + \cdots $. 
The summation over $r\in \bz_2$ 
then projects the KK momenta onto $n \in 2\bz+1$,
and the conformal weights read 
\begin{equation}
h_L = \frac{1}{4} \left(\frac{n}{2R} \pm R \right)^2, \hspace{1cm}
h_R = \frac{1}{2} 
    + \frac{1}{4} \left(\frac{n}{2R} \mp R \right)^2, \hspace{1cm} (n\in 2\bz+1).
\end{equation}
Their minima for the physical states are achieved by setting $n=\pm 1$, to give 
\begin{equation}
h_L = h_R = \frac{1}{2} + \frac{1}{4} \left(\frac{1}{2R} -R \right)^2 \geq \frac{1}{2}.
\label{lightest 2}
\end{equation}
This means that the winding states from these sectors 
are always massive except at 
the special radius $R= \frac{1}{\sqrt{2}}$, where 
extra massless excitations appear.

\end{itemize}
 
These demonstrate that no winding tachyons emerge in  
the non-SUSY vacuum \eqn{non-SUSY vacuum}.

~

The non-SUSY vacua  associated with $F_{(*,*)}^{T^2[SO(4)] \times T^2[SO(4)]}$ in
\eqn{Fab T2 T2} and $\tF_{(*,*)}^{T^2[SO(4)] \times T^2[SO(4)]}$  in
\eqn{tFab T2 T2} can be examined in a parallel way, and we obtain 
almost the same spectra of the winding excitations. 
However,  there is a slight difference  for the sectors of 
$w\in 2\bz+1$ in the  model from $\tF_{(*,*)}^{T^2[SO(4)] \times T^2[SO(4)]}$. 
In this case, the conformal weights of the 
$w=\pm 1$ sectors become
\begin{equation}
h_L = \frac{1}{4} 
  + \frac{1}{4} \left(\frac{n}{2R} \pm R \right)^2, \hspace{1cm}
h_R = \frac{1}{2} 
 + \frac{1}{4} \left(\frac{n}{2R} \mp R \right)^2, \hspace{1cm} (n\in \bz+\frac{1}{2}).
\end{equation}
Here, $h_L$ also acquires the twisted energy 
from the extra $\th_2$-factor.
The KK momenta are shifted by one half due to the absence
of the phase factors depending on the temporal winding $m$ 
(see \eqn{tFab T2 T2}). Consequently,
the lightest excitations lie in the sectors with $w= \pm 1, n= \pm \frac{1}{2}$,  
giving
\begin{equation}
h_L = h_R = \frac{1}{2} + \frac{1}{4} \left(\frac{1}{4R} -R \right)^2 \geq \frac{1}{2}.
\label{lightest 3}
\end{equation}
Again these are always massive except at 
the massless point $R= \frac{1}{2}$.

Also, for the $F_{(*,*)}^{T^2[SO(4)] \times T^2[SO(4)]}$ model, 
we find that {\em both} of \eqn{lightest 2} and \eqn{lightest 3} emerge as light  excitations with $w=\pm 1$, which get massless  at 
$R=\frac{1}{\sqrt{2}}$ and $R=\frac{1}{2}$ respectively.


~


\section{Summary and Discussions}

In this paper, we have studied 
type II string vacua which are defined by the asymmetric orbifolding  
based on 
the chiral reflections/T-duality twists in $T^4$ combined with 
the shift in the base circle,
in such a way that the modular invariance is kept manifest. 
They represent the non-geometric string vacua for T-folds, 
supposed that the world-sheet description of T-folds is generally given by 
asymmetric/generalized orbifolds. 
Including appropriate phases as in (\ref{tFab T2 T2}), 
the full OPE also respects the invariance under the T-duality twists 
in accord with \cite{HW}.
As the main result,  we have presented  simple examples of the non-SUSY 
vacua with vanishing cosmological constant 
at one loop. 
We summarize the points to be emphasized as follows:


\begin{itemize}
\item Our non-SUSY vacuum \eqn{non-SUSY vacuum} has been 
  defined by a cyclic orbifold which is generated by a 
single element $g$ in \eqn{twist g}. 
Thus, it provides a simpler model than the previous ones 
 \cite{Kachru1,Kachru2,Kachru3,Harvey,Shiu-Tye,Blumenhagen:1998uf,Angelantonj:1999gm}.
In this construction, 
taking {\em both} the asymmetric  orbifold action with $\sigma^2 = (-1)^{F_R}$ 
and the Scherk-Schwarz compactification (orbifolding by $\left. (-1)^{F_L} \otimes \cT_{2\pi R}\right|_{\msc{base}}$)
at the same time is  truly crucial in order to make the SUSY-breaking compatible with the bose-fermi cancellation.
Indeed, it is important that 
the left and right-moving non-SUSY chiral blocks 
$\f_{(*,*)}(\tau)$, $\overline{\f_{(*,*)}(\tau)}$,  which originate from the SUSY-breaking twists $(-1)^{F_L}$, $(-1)^{F_R}$, 
 depend on the winding numbers along the Scherk-Schwarz circle  {\em in an asymmetric way\/}.

\item The modular invariant partition function given in \eqn{non-SUSY vacuum} is 
$q$-expanded so as to be compatible with 
unitarity, as shown in subsection \ref{unitarity}.
Curiously, it turns out that the {\em left-moving\/} bose-fermi 
cancellation occurs in the even winding sectors, while we have 
the  {\em right-moving\/} bose-fermi cancellation 
in the odd winding sectors. This aspect is in sharp contrast with 
any SUSY vacua.

\item Despite the absence of the space-time SUSY and adopting 
the Scherk-Schwarz type compactification, 
we are free from the tachyonic instability at any radius 
of the Scherk-Schwarz circle.

\end{itemize}



To conclude, 
we would like to make a few comments on possible future studies.
First of all, it would 
indeed be an interesting issue whether 
our non-SUSY vacuum \eqn{non-SUSY vacuum} has vanishing cosmological 
constant at higher loops. 
Since the orbifold structure of this vacuum is simpler than those
of the previous ones quoted above, 
it would be worthwhile  to examine especially the two-loop case  
by following the analysis  
in \cite{ADP}.

Secondly,  in order to search a more broad class of such vacua, 
one may extend the  construction in this paper 
to other toroidal models of asymmetric orbifolds.
Furthermore, toward more realistic models,
it would also be   important to consider the non-geometric string vacua 
from SCFTs other than the toroidal ones. 
For previous attempts based on the $\cN=2$ SCFTs, see {\em e.g.}  \cite{KawaiS2}.
A challenging direction in this respect, and along \cite{SatohS},
 would be to construct such vacua based on the generalized
orbifolds through the topological interfaces,
which are wrapped around the cycles of the world-sheet torus 
in correlation with the shift operators.%
\footnote 
 {For applications of the world-sheet conformal interfaces 
to string theory, see {\em e.g.}  
\cite{Bachas:2007td,Satoh:2011an,Bachas:2012bj,Elitzur:2013ut}}
The point here would be how to organize the world-sheet chiral sectors
depending on the winding numbers along the Scherk-Schwarz like circle,
so that the bose-fermi cancellation does occur. 
We expect that the novel feature of the cancellation, which is remarked
 at the end of subsection \ref{unitarity}, 
would  be observed 
only in the non-geometric backgrounds.

Thirdly, one may also extend this work  
so as to include the open string sectors, namely, D-branes. 
Possibilities of the bose-fermi cancellation in the  open string Hilbert space 
have been investigated \cite{GabS}
under particular SUSY breaking configurations of D-branes. 
Closely related studies of D-branes in 
asymmetric orbifolds by the T-duality twists  
have been presented 
{\em e.g.} in \cite{BRR,GabSch,KawaiS1,Bianchi}.
It  would be 
interesting to study the aspects of D-branes in the type II vacua given 
in this paper (and their variants),
in comparison with these previous works.

~


\section*{Acknowledgments}
We would like to thank K. Aoki for a useful conversation.
This work is supported in part by 
JSPS KAKENHI Grant Number 24540248 and 23540322
from Japan Society for the Promotion of Science (JSPS).


\newpage

\section*{Appendix A: ~ Summary of  Conventions and  Useful Formulas}

\setcounter{equation}{0}
\def\theequation{A.\arabic{equation}}

\noindent
\underline{\bf Theta functions:}
%
 \begin{align}
 & \dsp \th_1(\tau,z):=i\sum_{n=-\infty}^{\infty}(-1)^n q^{(n-1/2)^2/2} y^{n-1/2}
  \equiv  2 \sin(\pi z)q^{1/8}\prod_{m=1}^{\infty}
    (1-q^m)(1-yq^m)(1-y^{-1}q^m), \nn [-10pt]
   & \\[-5pt]
 & \dsp \th_2(\tau,z):=\sum_{n=-\infty}^{\infty} q^{(n-1/2)^2/2} y^{n-1/2}
  \equiv 2 \cos(\pi z)q^{1/8}\prod_{m=1}^{\infty}
    (1-q^m)(1+yq^m)(1+y^{-1}q^m), \\
 & \dsp \th_3(\tau,z):=\sum_{n=-\infty}^{\infty} q^{n^2/2} y^{n}
  \equiv \prod_{m=1}^{\infty}
    (1-q^m)(1+yq^{m-1/2})(1+y^{-1}q^{m-1/2}),  
\\
 &  \dsp \th_4(\tau,z):=\sum_{n=-\infty}^{\infty}(-1)^n q^{n^2/2} y^{n}
  \equiv \prod_{m=1}^{\infty}
    (1-q^m)(1-yq^{m-1/2})(1-y^{-1}q^{m-1/2}) . 
 \end{align}
\begin{eqnarray}
 \Th{m}{k}(\tau,z)&:=&\sum_{n=-\infty}^{\infty}
 q^{k(n+\frac{m}{2k})^2}y^{k(n+\frac{m}{2k})} ,
\\
\tTh{m}{k}(\tau,z)&:=&\sum_{n=-\infty}^{\infty}
 (-1)^n q^{k(n+\frac{m}{2k})^2}y^{k(n+\frac{m}{2k})} ,
\\
\eta(\tau)  &:=& q^{1/24}\prod_{n=1}^{\infty}(1-q^n).
 \end{eqnarray}
 Here, we have set $q:= e^{2\pi i \tau}$, $y:=e^{2\pi i z}$  
 ($\any \tau \in \bh^+$, $\any z \in \bc$),
 and used abbreviations, $\th_i (\tau) \equiv \th_i(\tau, 0)$
 ($\th_1(\tau)\equiv 0$), 
$\Th{m}{k}(\tau) \equiv \Th{m}{k}(\tau,0)$.
It is straightforward to prove the following identities:
\begin{eqnarray}
 && \frac{\tTh{0}{1}(\tau)}{\eta(\tau)}=
 \sqrt{\frac{2\eta(\tau)}{\th_2(\tau)}}~ ,~~~
\frac{\Th{1/2}{1}(\tau)}{\eta(\tau)}=
 \sqrt{\frac{\eta(\tau)}{\th_4(\tau)}}~, ~~~
\frac{\tTh{1/2}{1}(\tau)}{\eta(\tau)}=
 \sqrt{\frac{\eta(\tau)}{\th_3(\tau)}}~.
\label{theta identity}
\end{eqnarray}

~


\noindent
\underline{\bf Poisson resummation formula:}
\begin{eqnarray}
&& \sum_{n\in\bsz}\exp\left(-\pi \al (n+a)^2+2\pi i b (n+a)\right)
=\frac{1}{\sqrt{\al}}\sum_{m\in\bsz}\exp
\left(-\frac{\pi(m-b)^2}{\al}+2\pi i m a\right), 
\nn
&& \hspace{10cm}
(\al >0 , ~ a,b \in \br).
\label{PR formula}
\end{eqnarray}

~


\section*{Appendix B: ~ Summary of Building Blocks}

\setcounter{equation}{0}
\def\theequation{B.\arabic{equation}}

In Appendix B, 
we summarize the notations of relevant building blocks
to construct the torus partition functions used in the main text.

~

\noindent
\underline{\bf Bulidng Blocks for the 
Bosonic $T^4$-secotor:}

\begin{description}
\item[1.  Chiral reflection in {$T^4[SO(8)]$}: ]
\begin{eqnarray}
&& 
\hspace{-1.5cm}
F^{T^4[SO(8)]}_{(a,b)}(\tau,\bar{\tau}) =
\left\{
\begin{array}{ll}
(-1)^{\frac{a}{2}}
 \overline{\left(\frac{\th_3\th_4}{\eta^2}\right)^2}
\cdot 
\frac{1}{2}\left\{
\left(\frac{\th_3}{\eta}\right)^4
+\left(
\frac{\th_4}{\eta}
\right)^4
\right\}
& ~~ (a\in 2\bz, ~ b\in 2\bz+1), \\
(-1)^{\frac{b}{2}}
 \overline{\left(\frac{\th_2\th_3}{\eta^2}\right)^2}
\cdot 
\frac{1}{2}\left\{
\left(\frac{\th_3}{\eta}\right)^4+\left(
\frac{\th_2}{\eta}
\right)^4
\right\}
& ~~ (a\in 2\bz+1, ~ b\in 2\bz), \\
 e^{-\frac{i\pi}{2}ab} 
 \overline{\left(\frac{\th_4\th_2}{\eta^2}\right)^2}
\cdot 
\frac{1}{2}\left\{
\left(\frac{\th_4}{\eta}\right)^4- \left(
\frac{\th_2}{\eta}
\right)^4
\right\}
& ~~ (a\in 2\bz+1, ~ b\in 2\bz+1), \\
\frac{1}{2}\left\{
\left|\frac{\th_3}{\eta}\right|^8 + \left|\frac{\th_4}{\eta}\right|^8
+\left|\frac{\th_2}{\eta}\right|^8
\right\}
& ~~ (a \in 2\bz, ~ b\in 2\bz) .
\end{array}
\right. 
\label{Fab T4 app}
\end{eqnarray}
Of course, 
$F^{T^4[SO(8)]}_{(a,b)}$ for the $a, b \in 2\bz$ case coincides with the original partition function 
$Z_{T^4[SO(8)]}$ \eqn{Z T4 SO(8)}.


\item[2.  Chiral reflection in {$T^2[SO(4)] \times T^2[SO(4)]$}: ]
%
\begin{eqnarray}
&& 
\hspace{-1.5cm}
F^{T^2[SO(4)]\times T^2[SO(4)]}_{(a,b)}(\tau,\bar{\tau}) 
\nn
&& 
=
\left\{
\begin{array}{ll}
 (-1)^{\frac{a}{2}} \overline{\left(\frac{\th_3\th_4}{\eta^2}\right)^2}
\cdot 
\frac{1}{4}\left\{
\left(\frac{\th_3}{\eta}\right)^2+(-1)^{\frac{a}{2}}\left(
\frac{\th_4}{\eta}
\right)^2
\right\}^2
& ~~ (a\in 2\bz, ~ b\in 2\bz+1), \\
 (-1)^{\frac{b}{2}}\overline{\left(\frac{\th_2\th_3}{\eta^2}\right)^2}
\cdot 
\frac{1}{4}\left\{
\left(\frac{\th_3}{\eta}\right)^2+(-1)^{\frac{b}{2}}\left(
\frac{\th_2}{\eta}
\right)^2
\right\}^2
& ~~ (a\in 2\bz+1, ~ b\in 2\bz), \\
 e^{-\frac{i\pi}{2}ab} \overline{\left(\frac{\th_4\th_2}{\eta^2}\right)^2}
\cdot 
\frac{1}{4}\left\{
\left(\frac{\th_4}{\eta}\right)^2-i(-1)^{\frac{a+b}{2}}\left(
\frac{\th_2}{\eta}
\right)^2
\right\}^2
& ~~ (a\in 2\bz+1, ~ b\in 2\bz+1), \\
\frac{1}{4}\left\{
\left|\frac{\th_3}{\eta}\right|^4+ \left|\frac{\th_4}{\eta}\right|^4
+\left|\frac{\th_2}{\eta}\right|^4 
\right\}^2
& ~~ (a \in 2\bz, ~ b\in 2\bz) .
\end{array}
\right. 
\label{Fab T2 T2}
\end{eqnarray}
Here, $F^{T^2[SO(4)]\times T^2[SO(4)]}_{(a,b)}$ for $a, b \in 2\bz$  coincides with the partition function 
$Z_{T^2[SO(4)]\times T^2[SO(4)]}$ \eqn{Z T4  2 SO(4)},
and we also find the identities
\begin{eqnarray}
\frac{1}{4}\left\{
\left|\frac{\th_3}{\eta}\right|^4+\left|\frac{\th_4}{\eta}\right|^4
+\left|\frac{\th_2}{\eta}\right|^4
\right\}^2
& = & 
\frac{1}{2} \left[ Z_{T^4[SO(8)]} (\tau, \bar{\tau}) 
+  \left|\frac{\th_3 \th_4}{\eta^2} \right|^4
+  \left|\frac{\th_4 \th_2}{\eta^2} \right|^4 
+  \left|\frac{\th_2 \th_3}{\eta^2} \right|^4
\right]
\nn
& = & \left[\left|\frac{\Th{0}{1}}{\eta} \right|^2
+ \left|\frac{\Th{1}{1}}{\eta} \right|^2
\right]^4 .
\label{Z T4 2 SO(4) app}
\end{eqnarray}
They obviously show the equivalence \eqn{T2 T2 rel}.


%
\item[3. Chiral reflection in {$T^2[SO(4)] \times T^2[SO(4)]$} with a phase factor: ]
\begin{eqnarray}
&& 
\hspace{-1.5cm}
\tF^{T^2[SO(4)]\times T^2[SO(4)]}_{(a,b)}(\tau,\bar{\tau}) 
= 
\left\{
\begin{array}{ll}
\left|\frac{\th_3\th_4}{\eta^2} \right|^4
& ~~ (a\in 2\bz, ~ b\in 2\bz+1), \\
\left|\frac{\th_2\th_3}{\eta^2} \right|^4
& ~~ (a\in 2\bz+1, ~ b\in 2\bz), \\
\left|\frac{\th_4\th_2}{\eta^2} \right|^4 
& ~~ (a\in 2\bz+1, ~ b\in 2\bz+1), \\
\frac{1}{4}\left\{
\left|\frac{\th_3}{\eta}\right|^4+ \left|\frac{\th_4}{\eta}\right|^4
+\left|\frac{\th_2}{\eta}\right|^4 
\right\}^2
& ~~ (a \in 2\bz, ~ b\in 2\bz) .
\end{array}
\right.
\label{tFab T2 T2}
\end{eqnarray}
Again the building block for $a, b \in 2\bz$ coincides with \eqn{Z T4 2 SO(4)},
and the blocks for the $(0, b)$-sectors with $b \in 2\bz+1$
are explicitly computed as
\begin{eqnarray}
\tF^{T^2[SO(4)]\times T^2[SO(4)]}_{(0,b)}(\tau,\bar{\tau}) 
& = & \left[ \, \overline{\sqrt{\frac{2\eta}{\th_2}}} 
\cdot \frac{1}{\eta} 
\sum_{n\in \bz}\, (-1)^n q^{n^2} \right]^4
\nn
& = & \left|\frac{\th_3\th_4}{\eta^2} \right|^4.
\end{eqnarray} 
Here, we used the identity  \eqn{theta identity} and the Euler identity 
$\th_2 \th_3 \th_4 = 2 \eta^3$ 
to derive the second line.
The building blocks of other sectors are obtained by requiring the modular covariance,
and one can quickly reproduce the results \eqn{tFab T2 T2}.

\end{description}

~




\noindent
\underline{\bf Chiral Building Blocks for the Fermionic Sector}

\begin{description}
%
%
\item[(i) For the case {$\sigma_L^2= {\bf 1}$}:]
\begin{eqnarray}
 f_{(a,b)}(\tau)
& =  & 2 q^{\frac{1}{4}a^2}e^{\frac{i\pi}{2}ab}
\,
\left(\frac{\th_1\left(\tau,\frac{a\tau+b}{2}\right)}{\eta(\tau)}\right)^2
\left(\frac{\th_1(\tau,0)}{\eta(\tau)}\right)^2 
\nn
&\equiv &
\left\{
\begin{array}{ll}
(-1)^{\frac{a}{2}}
\left\{
\left(\frac{\th_3}{\eta}\right)^2\left(\frac{\th_4}{\eta}\right)^2
- \left(\frac{\th_4}{\eta}\right)^2\left(\frac{\th_3}{\eta}\right)^2
+0 \right\}
 &  ~~ (a\in 2\bz,~ b\in 2\bz+1) ,\\
(-1)^{\frac{b}{2}}
\left\{
\left(\frac{\th_3}{\eta}\right)^2\left(\frac{\th_2}{\eta}\right)^2 +0 
- \left(\frac{\th_2}{\eta}\right)^2\left(\frac{\th_3}{\eta}\right)^2
\right\}
 &  ~~ (a\in 2\bz+1,~ b\in 2\bz) ,\\
-  e^{\frac{i\pi}{2}ab}
\left\{ 0+
\left(\frac{\th_2}{\eta}\right)^2\left(\frac{\th_4}{\eta}\right)^2
- \left(\frac{\th_4}{\eta}\right)^2\left(\frac{\th_2}{\eta}\right)^2
\right\}
 &  ~~ (a\in 2\bz+1,~ b\in 2\bz+1) , \\
\left(\frac{\th_3}{\eta}\right)^4
- \left(\frac{\th_4}{\eta}\right)^4
-\left(\frac{\th_2}{\eta}\right)^4
 & ~~ (a \in 2\bz, ~ b\in 2\bz) .
\end{array}
\right.
\label{fab app}
\end{eqnarray}


Since $f_{(a,b)} (\tau)$ actually vanish, 
it is better to first consider 
\begin{eqnarray}
 && f_{(a,b)}(\tau,\ep) \equiv 
2 q^{\frac{1}{4}a^2}e^{\frac{i\pi}{2}ab}
\,
\left(\frac{\th_1\left(\tau,\frac{a\tau+b}{2}\right)}{\eta(\tau)}\right)^2
\left(\frac{\th_1(\tau,\ep)}{\eta(\tau)}\right)^2 ,
\end{eqnarray}
in order to express the modular covariance relation with no subtlety, 
One should then 
interpret \eqn{mc fab} as the $\ep\,\rightarrow \, 0$ limit of  
\begin{eqnarray}
 && f_{a,b}(\tau, \ep)|_S \equiv f_{(a,b)}\left(-\frac{1}{\tau}, 
\frac{\ep}{\tau}\right)
= e^{i\pi \frac{2}{\tau}\ep^2}
f_{(b,-a)}(\tau, \ep), \nn
&& f_{(a,b)}(\tau,\ep)|_T \equiv f_{(a,b)}(\tau+1,\ep) 
= -e^{-2\pi i \frac{1}{6}} f_{(a,a+b)}(\tau,\ep) .
\label{mc fab refined}
\end{eqnarray}


\item[(ii) For the case of {$\sigma_L^2= (-1)^{F_L}$}: ]
\begin{eqnarray}
 && \f_{(a,b)}(\tau) = f_{(a,b)}(\tau), ~~~ 
(a\in 2\bz+1~  \mbox{or}~
 b \in 2\bz+1), 
 \label{fab 2 app}
\end{eqnarray}
and 
\begin{equation}
 \f_{(a,b)}(\tau) = \left\{
\begin{array}{ll}
\left(\frac{\th_3}{\eta}\right)^4
- \left(\frac{\th_4}{\eta}\right)^4
- \left(\frac{\th_2}{\eta}\right)^4
& ~~ (a\in 4\bz, ~ b \in 4\bz) , \\
\left(\frac{\th_3}{\eta}\right)^4
- \left(\frac{\th_4}{\eta}\right)^4
+ \left(\frac{\th_2}{\eta}\right)^4
& ~~ (a\in 4\bz, ~ b \in 4\bz+2) , \\
\left(\frac{\th_3}{\eta}\right)^4
+ \left(\frac{\th_4}{\eta}\right)^4
- \left(\frac{\th_2}{\eta}\right)^4
& ~~ (a\in 4\bz+2, ~ b \in 4\bz) , \\
-\left\{\left(\frac{\th_3}{\eta}\right)^4
+ \left(\frac{\th_4}{\eta}\right)^4
+ \left(\frac{\th_2}{\eta}\right)^4 \right\}
& ~~ (a\in 4\bz+2, ~ b \in 4\bz+2). \\
\end{array}
\right.
\label{fab 2-2 app} 
\end{equation}

\end{description}


\end{document}